\documentclass[%
reprint,
 amsmath,amssymb,
 aps,
 prl,
floatfix,
]{revtex4-2}

\usepackage{graphicx}
\usepackage{dcolumn}
\usepackage{bm}
\usepackage{CJK}

\usepackage{xcolor}
\usepackage{tikz}
\usetikzlibrary{shapes}
\definecolor{blue}{RGB}{0,112,192}
\definecolor{lightblue}{RGB}{0,176,240}
\definecolor{green}{RGB}{0,176,80}
\definecolor{yellow}{RGB}{255,255,0}
\definecolor{orange}{RGB}{255,192,0}
\definecolor{red}{RGB}{255,0,0}
\definecolor{darkred}{RGB}{118,0,0}
\definecolor{purple}{RGB}{208,0,154}

\newcommand*{\tikzcirc}[2]{%
   \setbox0=\hbox{\strut}%
   \begin{tikzpicture}
     \useasboundingbox (-.25em,0) rectangle (.25em,\ht0);
     \filldraw[draw=#1,fill=#2] (0.0,0.35\ht0) circle[radius=.25em];
   \end{tikzpicture}%
}
\newcommand*{\tikzrect}[2]{%
   \setbox0=\hbox{\strut}%
   \begin{tikzpicture}
     \useasboundingbox (-.25em,0) rectangle (.25em,\ht0);
     \filldraw[draw=#1,fill=#2] (-0.25em,0.05em) rectangle (.25em,0.55em);
   \end{tikzpicture}%
}
\newcommand*{\tikzdiam}[2]{%
   \setbox0=\hbox{\strut}%
   \begin{tikzpicture}[rotate = 45]
     \useasboundingbox (-.25em,0) rectangle (.25em,\ht0);
     \filldraw[draw=#1,fill=#2] (-0.25em,0.0em) rectangle (.25em,0.5em);  
   \end{tikzpicture}%
   }
   
\newcommand{\tikztri}[2]{\tikz{\node[draw=#1,fill=#2,isosceles triangle,isosceles triangle stretches,shape border rotate=90,minimum width=0.2cm,minimum height=0.2cm,inner sep=0pt] at (0,0) {};}}

\begin{document}
\preprint{APS/123-QED}

\title{Gyratory Shearing Compaction of Granular Materials}

\author{Teng Man}
 \email{manteng@westlake.edu.cn}
 \email{manxx027@umn.edu}
 \altaffiliation[Work partly done at the ]{%
 Department of CEGE, UMN, MN, USA}%
 \affiliation{School of Engineering, Westlake University, Hangzhou, Zhejiang 310024, China}


\date{\today}

\begin{abstract}
In this letter, instead of investigating the compaction of granular materials under tapping or cyclic shearing, we focused more on the gyratory shearing compaction, where particles are subjected to constant pressure and shear rate continuously. Such a phenomenon is crucial to the compaction of asphalt mixtures or soil in civil engineering, and can be extended to other areas, such as powder processing and pharmaceutical engineering. In this study, we found that the gyratory speed or interstitial fluid viscosity has almost no impact on the compaction behavior, while the pressure plays a more important role. Additionally, it is the inertial time scale which dictates the compaction behavior under gyratory shearing in most cases, whereas the viscous time scale can also have influence in some conditions.

\begin{description}
\item[Usage]
Secondary publications and information retrieval purposes.
\item[PACS numbers] 47.57.Gc, 81.05.Rm, 83.10.Pp, 83.80.Hj
\item[Structure] You may use the \textbf{description} environment to structure your abstract
\end{description}
\end{abstract}

\maketitle



Granular materials are commonly used in engineering practices such as pharmaceutical engineering, food processing, civil engineering, etc.. Granular materials in civil engineering are often subjected to complex loading conditions or loading path, or have complex particle shapes. One kind of loading condition, which fascinates the author, is the compaction of granular materials (or granular-fluid systems). In most cases, we would like to have our concrete or asphalt pavement with as minimum amount of air voids as possible. To achieve the compaction under both pressure and continuous shearing, a so-called gyratory compactor is often used in pavement industry to achieve denser asphalt mixtures. However, we know very little about the physics behind this kind of compaction when particles are subjected to continuous gyratory shearing (different from cyclic shearing where particles dilate and relax due to the cyclic excitation, i.e. the dilation/relaxation cycle).

Most previous research on granular compaction has focused on mono-disperse spherical particle systems without interstitial fluids (dry granular material) under tapping, vibrating or cyclic shearing. Mehta et al. \cite{mehta1990phenomenological,mehta1994dynamics} proposed a two-exponential equation with two relaxation time scales to describe the compaction behaviors based on experimental results \cite{knight1995density}.
\begin{equation}
    \begin{split}
        \phi(t) = a_0 - a_1 \textrm{exp}(-a_{3}t) - a_4\textrm{exp}(-a_{5}t) \label{eq1}
    \end{split}
\end{equation}
where $\phi(t)$ is the solid fraction of granular materials at time $t$, and $a_0$, $a_1$, $a_2$, $a_3$, $a_4$, $a_5$ are fitting parameters obtained from experimental data. The two exponential terms can be associated to the independent and collective motion of particles during the compaction process. Knight et al. \cite{knight1995density,nowak1998density} proposed a logarithmic equation with the following form (Eq.\ \ref{eq2}) to better fit the compaction behavior of granular materials.
\begin{equation}
\begin{split}
\phi(t) = \phi_{f} - \frac{\Delta \phi_{\infty}}{1+B\ln{(1+t/\tau)}} \label{eq2}
\end{split}
\end{equation}
where $\phi_f$, $\Delta\phi_{\infty}$, $B$, and $\tau$ are fitting parameters. Meanwhile, the Renne group found that the compaction curve can also be fitted using a stretched exponential curve as follows \cite{philippe2002compaction,philippe2003granular, richard2005slow}:
\begin{equation}
    \begin{split}
        \phi(t) = \phi_f - (\phi_f - \phi_0)e^{-(t/\tau)^\beta} \label{eq3}
    \end{split}
\end{equation}

In all these cases, the reason why granular systems can evolve to a more dense configuration is partly due to the dilation-relaxation cycle during the loading process. For example, during the tapping excitation, particles will first dilate and then find their lowest-energy-spot during the relaxation process. However, such a phenomenon does not exist during the gyratory compaction (the boundary condition of a gyratory compactor will be shown later in this letter), where the particles are continuously subjected to constant load continuously. So it is difficult to foresee the hidden time scale inside this problem, thus, difficult to quantify the compaction behavior of a gyratory shearing process.

In this letter, the author will first introduce this problem and the boundary condition of the gyratory compaction. Then, the computational simulations using the discrete element method (DEM) \cite{cundall1979discrete,tsuji1992} will be implemented to conduct a simple parametric study of the gyratory compaction of monodisperse granular assemblies. According to the simulation results, we can draw conclusions about the nature of the gyratory compaction of granular materials and provide some insights in the time scale governing the compaction speed.


Gyratory compaction is a type of compaction method commonly used in pavement engineering to achieve more dense asphalt mixture. During a gyratory compaction, the material is subjected to both constant pressure and gyratory motion. Figure 1 shows the configuration of a gyratory compactor, where particles are put inside a cylindrical "ring", and bounded by a top plate and a bottom plate. The top plate remains stationary in the vertical direction, while the bottom plate could move upward while maintaining a constant pressure, $\sigma_n$. The cylindrical ring is subjected to a gyratory motion with gyratory speed $n$ (SI unit, rps or rpm) and gyratory angle $\theta$. The gyratory motion can give the granular system an average shear rate equal to $\dot{\gamma} = 2\pi\theta n$.
\begin{figure}[!ht]
  \centering
  \includegraphics[scale = 0.5]{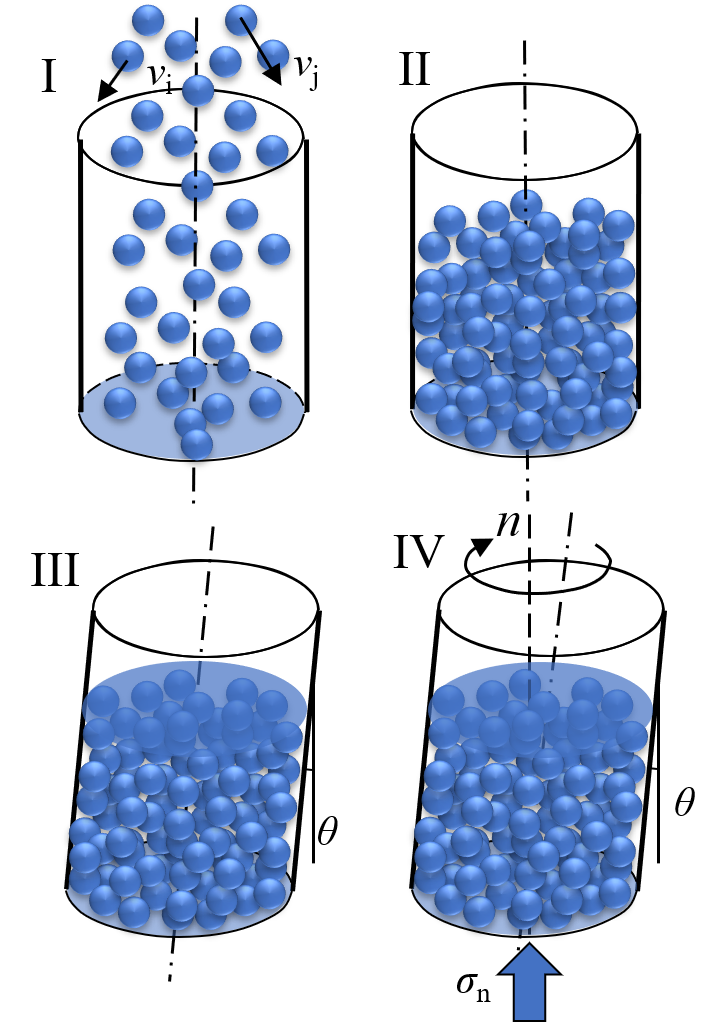}
  \caption{Configuration of the gyratory compactor and the model setup in the DEM simulation}
  \label{fig1}
\end{figure}

The setup of DEM simulations is the same as that of experiments shown in Fig. \ref{fig1}. We can see that the particles are firstly dropped randomly (Fig. 1I) into a cylindrical box to form a random packing (Fig. 1II). Then, in the simulation, we slightly tilte the cylinder to provide a gyratory angle of $\theta$, while adding a top plate to the cylindrical box. Finally, the loosely packed granular assemblies are loaded and sheared with constant pressure and constant gyratory speed (Fig. 1IV). In our model, two kinds of particle interactions is considered. The first type of interaction considered is the physical contact among particles. When two particles have physical contact, the interaction between them can be calculated based on Hertz-Mindlin contact model \cite{hill2014segregation,hill2011rheology}, where the normal and tangential contact forces are calculated using following equations.
\begin{subequations} \label{eq4}
\begin{align}
    F_{c}^{\rm{ij,n}} = -k_n\delta_{n}^{1.5} - \eta_n\delta_{n}^{0.25}\dot{\delta}_{n} \label{eq4a}\\
    F_{c}^{\rm{ij,t}} = \rm{min}\big( -k_{t}\delta_{n}^{0.5}\delta_t - \eta_{t}\delta_n^{0.25}\dot{\delta}_t, \mu_{p}\it{F}_{c}^{\rm{ij,n}} \big) \label{eq4b}
\end{align}
\end{subequations}
where $F_{c}^{\rm{ij,n}}$ and $F_{c}^{\rm{ij,t}}$ are normal and tangential contact forces acting on particle \textit{i} from particle \textit{j}. $\delta_{n}$ is the overlap between particles in normal direction in DEM simulation, which is given by $\delta_{n} = R_{i}+R_{j} - |\vec{r}_{i} - \vec{r}_{j}|$, where $R_i$ and $R_j$ are particle radii, and $\vec{r}_{i}$ and $\vec{r}_{j}$ are position vectors of two particles. $\delta_t$ is the corresponding tangential deformation at the contact point between two particles, and $\mu_{p}$ is the coefficient of friction between particles. The second type of particle interactions is considered when the interstitial viscous fluid exists among particles, where the lubrication effect can be calculated accordingly \cite{trulsson2012transition,man2018rheology,man2019rheology,pitois2000,goldman1967}.
\begin{subequations} \label{eq5}
\begin{align}
F_{v}^{\rm{ij,n}} = 6\pi\eta_f R_{\rm{eff}}^{2}\frac{\dot{\delta}_g}{\delta_g} \label{eq5a}\\
F_{v}^{\rm{ij,t}} = 6\pi\eta_f R_{\rm{eff}}v_{t}^{\rm{rel}}\big[\frac{8}{15}\rm{ln}(R_{\rm{eff}}/\delta_{g})+0.9588\big] \label{eq5b}
\end{align}
\end{subequations}
where $F_{v}^{\rm{ij,n}}$ and $F_{v}^{\rm{ij,t}}$ are normal and tangential lubrication forces between particle $i$ and particle $j$. $\eta_f$ is the interstitial fluid viscosity and $R_{\rm{eff}}$ is the effective radius calculated based on the radius of two contacting particles. $v_{t}^{\rm{rel}}$ is the relative tangential velocity. $\delta_g$ is the gap between the nearest surface of two particles. The lubrication forces will diminish once the distance between particles exceeds the average particle diameters, $\delta_{\rm{max}} = \bar{d}$, and will remain invariant to distance once the particle distance is smaller than $\delta_{min} = 0.1\bar{d}$, which represents the roughness of the particles. No global drag force is considered in our simulation, since we focus more on systems where the fluid can travel together with the particles. We assume that there is no relative motion between interstitial fluid and individual particles. Additionally, in this work, we focus on granular-fluid systems which exceed the pendulum or capillary stage, so that we do not need to consider the effect of capillary forces between particles (this is also similar to that of asphalt mixtures where viscous forces play a more important role.).

We calculated the parameters of the particle interaction law using the same method as that in Ref. \cite{hill2014segregation}. The elastic modulus of particles is 30 GPa, the Poisson ratio is equal to 0.2, the density of the particles is 2650 kg/m$^{2}$, and the frictional coefficient between particles is set to be 0.2. The size of the particles is uniformly distributed between $0.8\bar{d}$ and $1.2\bar{d}$, where $\bar{d} = 2.0$ mm. The diameter of the cylindrical ring was set equal to 20 times the average size of the particles. The gyratory angle, $\theta$, is kept at 0.25$^{\circ}$, which is similar to that used in the gyratory compaction of asphalt mixtures. In this study, we varied the pressure from 10 kPa to 1000 kPa, the gyratory speed from 3 rpm to 100 rpm, and the viscosity of the interstitial fluid from 0 cP (no interstitial fluid) to 1000 cP (same as that of glycerin).



\begin{figure}[!ht]
  \centering
  \includegraphics[scale = 0.35]{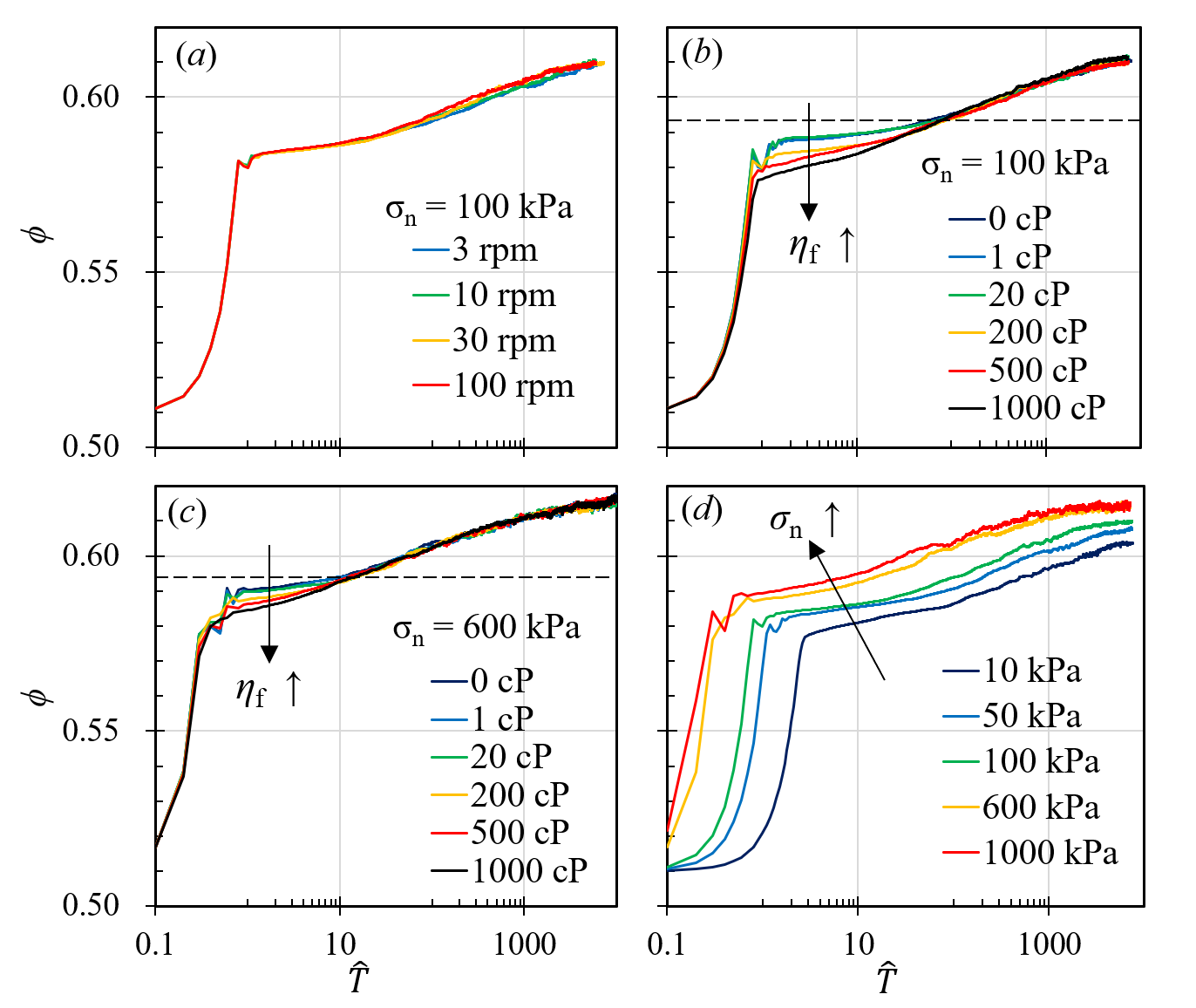}
  \caption{Compaction curves of granular materials with different conditions. In (a), we plotted the results of simulations with different gyratory speeds while keeping the pressure and the interstitial fluid viscosity constant at 100 kPa and 100 cP, respectively. We plotted the simulations with different interstitial fluid viscosity in (b) and (c). However, the pressure is different in these two sets of simulations. In (d), we have the relationship between solid fraction and dimensionless time, $\hat{T}$, of simulations with different pressures.}
  \label{fig2}
\end{figure}

Figure 2(a) shows compaction curves of granular materials with different gyratory speeds while keeping the pressure and interstitial fluid viscosity constant at 100 kPa and 100 cP, respectively, by plotting the relationship between solid fraction $\phi$ and normalized time $\hat{T} = t\sqrt{g/\bar{r}_p}$, where $\bar{r}_p = 0.5\bar{d}$ is the average particle radius. The gyratory compaction of granular materials can be roughly classified into two stages: (i) at the beginning of the compaction, the solid fraction increased rapidly from loosely random packing solid fraction to random packing fraction; (ii) Afterwards, the solid fraction exceeds the random packing fraction and has logarithmic growth with respect to time. Between these two stages, there is a seemingly intermediate stage where the slope of the curve is slightly smaller than that in the logarithmic growth region. 

In previous research, when granular systems were under cyclic shearing the intensity of the shearing would influence dramatically the compaction behavior. However, in our research, as shown in Fig. 2(a), changing the gyratory speed (which can change the average shear rate) does not have a large effect on the overall compaction behavior. The change of gyratory speed has no influence on the statistical behavior of the granular materials under gyratory speed, either. We also investigated the influence of interstitial fluid viscosity. Fig. 2(b) and (c) show results of two sets of simulations. For one set of simulation (Fig. 2(b)), we kept the pressure and gyratory speed at 100 kPa and 30 rpm, while changing the viscosity from 0 cP to 1000 cP. Intuitively, we would expect that the viscosity of interstitial fluid would have a huge impact on the behavior of compaction, which was shown in the work of Fiscina et al \cite{fiscina2010compaction}. Our work shows that the existence of interstitial fluid only influences the gyratory compaction of granular materials at the intermediate stage. One possible reason which leads us to different results might be the fact that we have neglected capillary forces.

\begin{figure}[!ht]
  \centering
  \includegraphics[scale = 0.35]{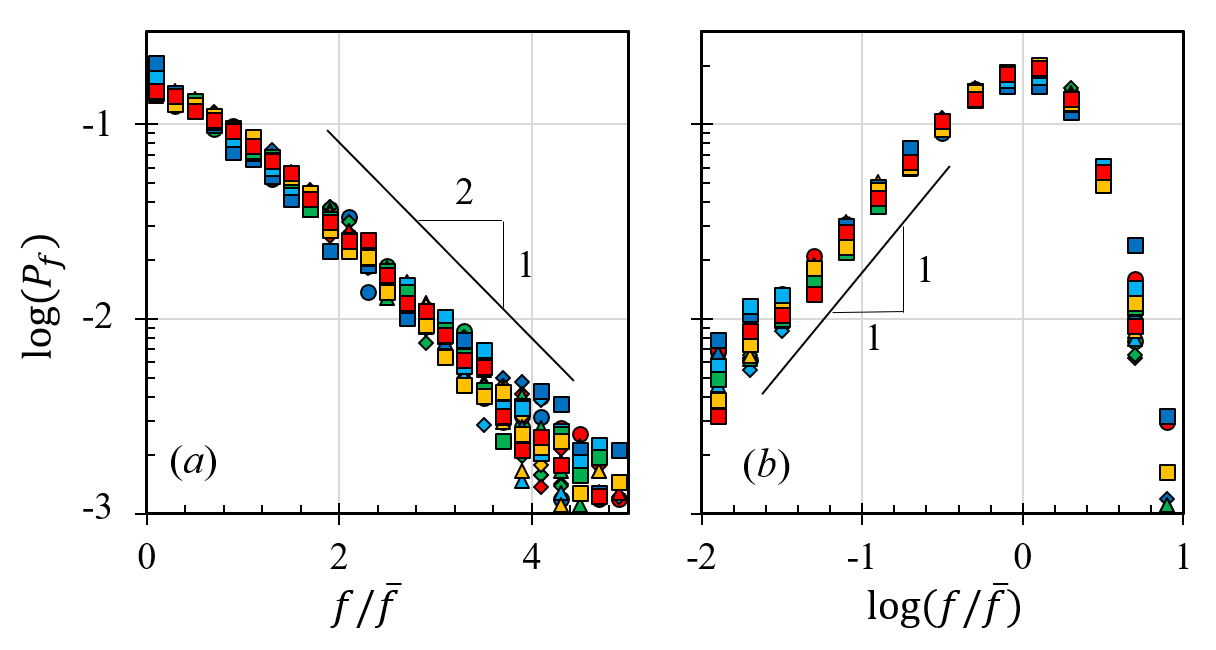}
  \caption{Histogram of the contact forces between particles. Markers \protect\tikzcirc{black}{blue}, \protect\tikzcirc{black}{green}, \protect\tikzcirc{black}{orange}, \protect\tikzcirc{black}{red} represent simulation results of different gyratory speeds. \protect\tikzdiam{black}{blue}, \protect\tikzdiam{black}{lightblue}, \protect\tikzdiam{black}{green}, \protect\tikzdiam{black}{yellow}, \protect\tikzdiam{black}{orange}, \protect\tikzdiam{black}{red} represent simulation results of different viscosity when $\sigma_n = 100$ kPa, while \protect\tikztri{black}{blue}, \protect\tikztri{black}{lightblue}, \protect\tikztri{black}{green}, \protect\tikztri{black}{yellow}, \protect\tikztri{black}{orange}, \protect\tikztri{black}{red} represent simulation results of different viscosity when $\sigma_n = 600$ kPa. \protect\tikzrect{black}{blue}, \protect\tikzrect{black}{lightblue}, \protect\tikzrect{black}{green}, \protect\tikzrect{black}{orange}, \protect\tikzrect{black}{red} represent simulation with different pressures}
  \label{fig3}
\end{figure}

In Figure 2(c), we plot the relationship between $\phi$ and $\hat{T}$ when the pressure is changed to 600 kPa. The time range, where the interstitial fluid has influence, decreases. In this case, the pressure is much higher which makes the granular assembly reach a certain solid fraction much earlier than that in Fig. 2(b). Interestingly, the solid fraction which marks the end of interstitial fluid interference is the same for different pressure conditions. When the solid fraction is larger than $\approx 0.59$, the viscosity of interstitial fluid has no effect on the gyratory compaction behavior of granular materials, no matter how large the viscosity becomes. Meanwhile, this dividing point for solid fraction happens to be equal to the loose random packing fraction \cite{dullien1992porous}. Also, Man and Hill \cite{man2018rheology,man2019rheology} found that, in a granular-fluid system, the frictional rheology (the $\mu_{\rm{eff}} - \phi$ relationship) followed two pathways depending on the magnitude of the viscous number $I_v = \eta_f\dot{\gamma}/\sigma_n$ in a shearing system, but these two different pathways merged into one curve at approximately $\phi = 0.59$ where fluid viscosity can no longer influence the rheology of granular materials. Our results in the gyratory compaction also conclude with a 0.59 solid fraction terminating the influence of interstitial fluid.

\begin{figure}[!ht]
  \centering
  \includegraphics[scale = 0.35]{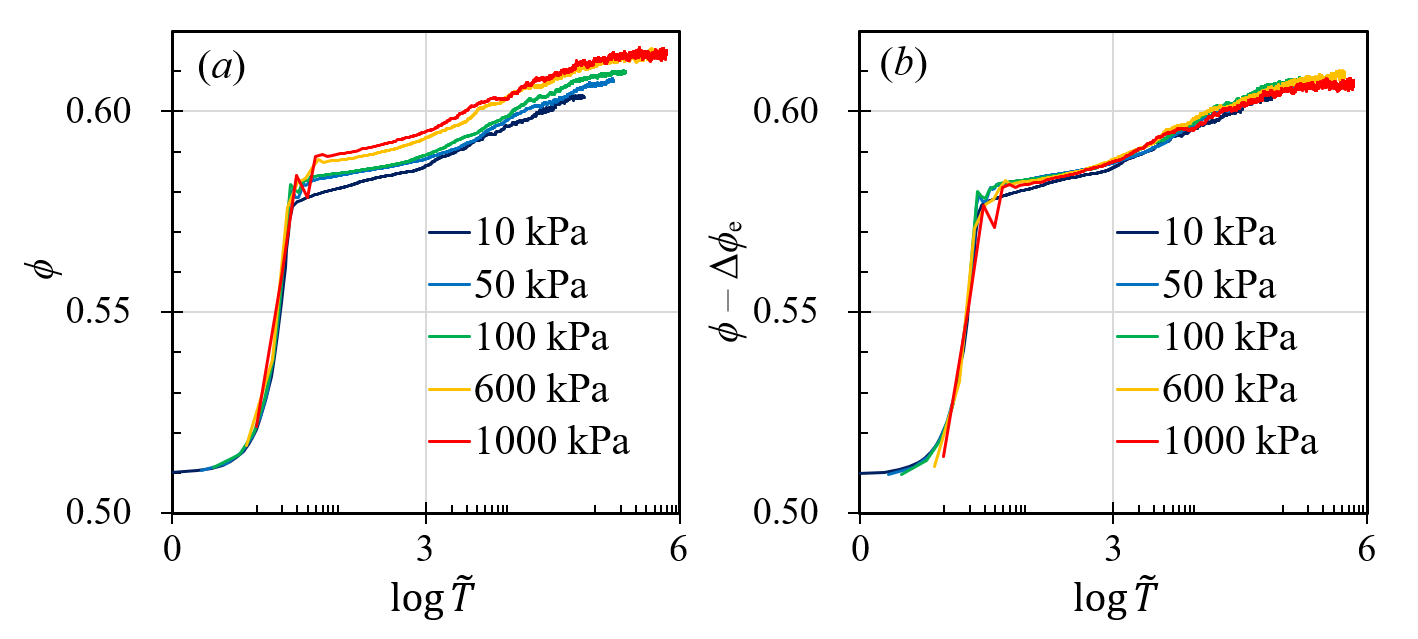}
  \caption{(a) Relationship between the solid fraction, $\phi$, and the dimensionless time, $\tilde{T} = t\sqrt{\sigma_n/\rho_p}/\bar{d}$. (b) Relationship between the "visco-plastic" solid fraction, $\phi-\Delta\phi_e$, and the dimensionless time, $\tilde{T}$.}
  \label{fig4}
\end{figure}

Different from the effect of gyratory speed and interstitial fluid viscosity, as we vary the pressure applied to the granular packing, the compaction behavior changes dramatically. Higher pressure could help granular materials achieve much higher solid fraction. It is also faster for higher pressure cases to obtain a certain solid fraction than low pressure simulations. Although the compaction behavior of systems with different pressure differs from each other, such changes of pressure do not influence the statistics of inter-particle forces much (Fig. 3). Fig. 3 shows that the statistics of the normalized contact forces always have an exponential decay when contact forces $f$ is larger than the mean contact forces $\bar{f}$, and a power law tail when contact forces is smaller than the mean contact forces. This leads to the question of what drives the change of compaction behavior, which is answered in this study. As what we have introduced, previous studies often link the compaction curve to some time scales without discussing the details of these time scales and the factors affecting them.. In our study, we could regard the compaction behavior as a competition among the shearing time scale, $\tau_s = 1/\dot{\gamma}$, the viscous time scale, $\tau_v = \eta_f/\sigma_n$, and the inertial time scale, $\tau_i = d/\sqrt{\sigma_n/\rho_p}$, as it is in the frictional rheology of granular materials \cite{jop2006constitutive,boyer2011unifying,man2018rheology}.

\begin{figure}[!ht]
  \centering
  \includegraphics[scale = 0.45]{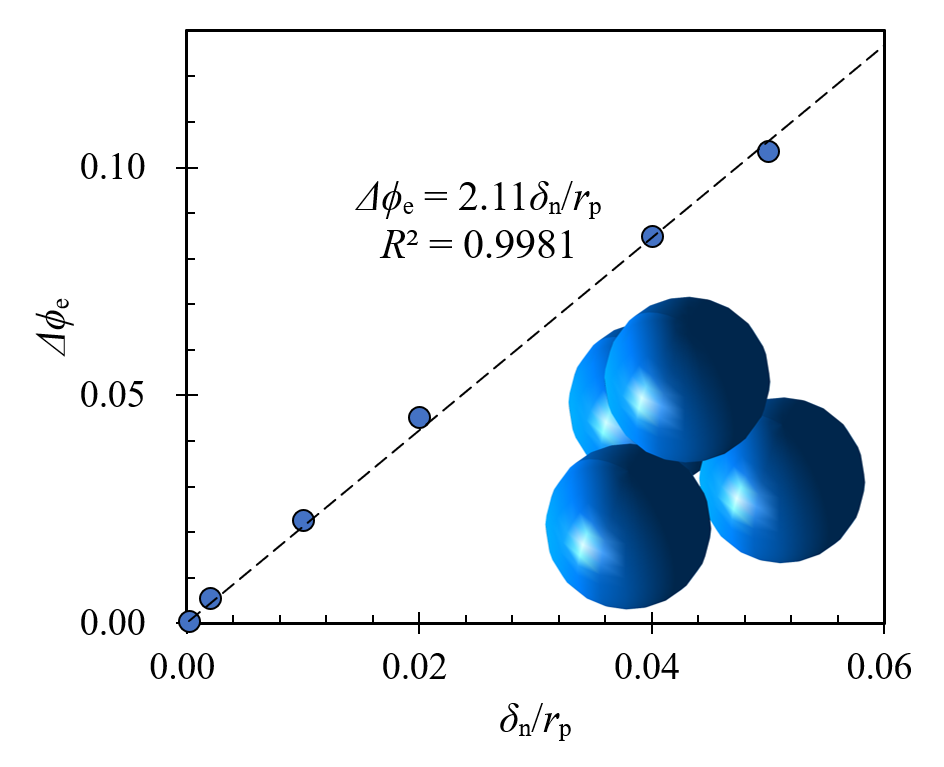}
  \caption{Relationship between the change of solid fraction, $\Delta\phi_e$, and the relative deformation between adjacent particles, $\delta_n/r_p$, where $r_p$ is the particle radius. The inset shows the basic structure of closed packing particles used for calculating the change of solid fractions.}
  \label{fig5}
\end{figure}

From Fig. 2(a) - (c), we can conclude that, comparing to the inertial effect, the influence of interstitial fluid viscosity and gyratory speed can be neglected. Thus, the compaction curve should be controlled solely by the inertial time scale, $\tau_i = d/\sqrt{\sigma_n/\rho_p}$. If we plot $\phi$ against the normalized time $\tilde{T} = t/\tau_i = t\sqrt{\sigma_n/\rho_p}/\bar{d}$, the compaction curves should collapse into one curve. However, as we can see from Fig. 4(a), the curves deviate from each other after entering the logarithmic stage. Interestingly, at same normalized time $\tilde{T}$, the slope of the curve is approximately the same for different pressure conditions, which make us believe that a factor is missing in this analysis. 

Thus, we divided the compaction into "elastic" compaction and "visco-plastic" compaction. The "elastic" compaction can be calculated based on the solid fraction change of granular packing with the basic hexagonal close packing (HCP) given pressure condition $\sigma_n$. In this case, the forces between adjacent particles are $F_c=0.25\sigma_n\pi d^2 = k_n\delta_n^{1.5}$, where $k_n = 1.33\sqrt{R_{\rm{eff}}}E_{\rm{eff}}$ according to Hertz contact model \cite{hill2014segregation}. Based on this, we can calculate the elastic deformation $\delta_n$. Using Monte Carlo algorithm, we could calculate the relationship between $\delta_n$ and the "elastic" change of solid fraction plotted in Fig. 5. With these analyses, we could link the pressure $\sigma_n$ to the change of solid fraction $\Delta\phi_e$. In Figure 4(b), we plotted the relationship between $\phi - \Delta\phi_e$ and $\tilde{T} = t\sqrt{\sigma_n/\rho_p}/\bar{d}$, which shows the good collapse of data among simulations with different pressures.

In this letter, the author analyzed the compaction behavior of granular materials during gyratory compaction, which is different from the compaction due to tapping or cyclic shearing. We concluded that the gyratory speed plays no effective role in influencing the gyratory compaction behavior, and that the viscosity of interstitial fluid only has secondary impact on the compaction behavior. It is the pressure which could greatly influence the compaction results. During the compaction, three different time scales, the shearing time scale, $1/\dot{\gamma}$, the viscous time scale, $\eta_f/\sigma_n$, and the inertial time scale, $d/\sqrt{\sigma_n/\rho_p}$, have to "fight against" each other for general dominance during the compaction process. Also, we could differentiate the deformation of granular materials during compaction into (i) "elastic" compaction and (ii) "visco-plastic" compaction. The "visco-plastic" compaction deformation ($\phi - \Delta\phi_e$) is controlled solely by the inertial time scale ,$d/\sqrt{\sigma_n/\rho_p}$, shown in Fig. 4(b). This research is particularly helpful to civil engineering industries where intensity and duration of compaction need to be determined beforehand for maximum economic gains. Also, such experiment can be beneficial to research understanding the force chains in 3D cases while implementing computed tomography technology in the future.

\begin{acknowledgments}
The author acknowledges the financial support from the CEGE Department (Sommerfeld Fellowship) and the China Center (Shaw-Lundquist Fellowship) at the University of Minnesota. The author would also like to thank Dr. Kimberly Hill and Anna Gorgogianni for the help during the discussion and writing of this letter.
\end{acknowledgments}




\bibliography{ManGyraComp}

\begin{thebibliography}{20}%
\makeatletter
\providecommand \@ifxundefined [1]{%
 \@ifx{#1\undefined}
}%
\providecommand \@ifnum [1]{%
 \ifnum #1\expandafter \@firstoftwo
 \else \expandafter \@secondoftwo
 \fi
}%
\providecommand \@ifx [1]{%
 \ifx #1\expandafter \@firstoftwo
 \else \expandafter \@secondoftwo
 \fi
}%
\providecommand \natexlab [1]{#1}%
\providecommand \enquote  [1]{``#1''}%
\providecommand \bibnamefont  [1]{#1}%
\providecommand \bibfnamefont [1]{#1}%
\providecommand \citenamefont [1]{#1}%
\providecommand \href@noop [0]{\@secondoftwo}%
\providecommand \href [0]{\begingroup \@sanitize@url \@href}%
\providecommand \@href[1]{\@@startlink{#1}\@@href}%
\providecommand \@@href[1]{\endgroup#1\@@endlink}%
\providecommand \@sanitize@url [0]{\catcode `\\12\catcode `\$12\catcode
  `\&12\catcode `\#12\catcode `\^12\catcode `\_12\catcode `\%12\relax}%
\providecommand \@@startlink[1]{}%
\providecommand \@@endlink[0]{}%
\providecommand \url  [0]{\begingroup\@sanitize@url \@url }%
\providecommand \@url [1]{\endgroup\@href {#1}{\urlprefix }}%
\providecommand \urlprefix  [0]{URL }%
\providecommand \Eprint [0]{\href }%
\providecommand \doibase [0]{https://doi.org/}%
\providecommand \selectlanguage [0]{\@gobble}%
\providecommand \bibinfo  [0]{\@secondoftwo}%
\providecommand \bibfield  [0]{\@secondoftwo}%
\providecommand \translation [1]{[#1]}%
\providecommand \BibitemOpen [0]{}%
\providecommand \bibitemStop [0]{}%
\providecommand \bibitemNoStop [0]{.\EOS\space}%
\providecommand \EOS [0]{\spacefactor3000\relax}%
\providecommand \BibitemShut  [1]{\csname bibitem#1\endcsname}%
\let\auto@bib@innerbib\@empty
\bibitem [{\citenamefont {Mehta}\ and\ \citenamefont
  {Edwards}(1990)}]{mehta1990phenomenological}%
  \BibitemOpen
  \bibfield  {author} {\bibinfo {author} {\bibfnamefont {A.}~\bibnamefont
  {Mehta}}\ and\ \bibinfo {author} {\bibfnamefont {S.}~\bibnamefont
  {Edwards}},\ }\bibfield  {title} {\bibinfo {title} {A phenomenological
  approach to relaxation in powders},\ }\href@noop {} {\bibfield  {journal}
  {\bibinfo  {journal} {Physica A}\ }\textbf {\bibinfo {volume} {168}},\
  \bibinfo {pages} {714} (\bibinfo {year} {1990})}\BibitemShut {NoStop}%
\bibitem [{\citenamefont {Mehta}\ and\ \citenamefont
  {Barker}(1994)}]{mehta1994dynamics}%
  \BibitemOpen
  \bibfield  {author} {\bibinfo {author} {\bibfnamefont {A.}~\bibnamefont
  {Mehta}}\ and\ \bibinfo {author} {\bibfnamefont {G.}~\bibnamefont {Barker}},\
  }\bibfield  {title} {\bibinfo {title} {The dynamics of sand},\ }\href@noop {}
  {\bibfield  {journal} {\bibinfo  {journal} {Rep Prog Phys}\ }\textbf
  {\bibinfo {volume} {57}},\ \bibinfo {pages} {383} (\bibinfo {year}
  {1994})}\BibitemShut {NoStop}%
\bibitem [{\citenamefont {Knight}\ \emph {et~al.}(1995)\citenamefont {Knight},
  \citenamefont {Fandrich}, \citenamefont {Lau}, \citenamefont {Jaeger},\ and\
  \citenamefont {Nagel}}]{knight1995density}%
  \BibitemOpen
  \bibfield  {author} {\bibinfo {author} {\bibfnamefont {J.~B.}\ \bibnamefont
  {Knight}}, \bibinfo {author} {\bibfnamefont {C.~G.}\ \bibnamefont
  {Fandrich}}, \bibinfo {author} {\bibfnamefont {C.~N.}\ \bibnamefont {Lau}},
  \bibinfo {author} {\bibfnamefont {H.~M.}\ \bibnamefont {Jaeger}},\ and\
  \bibinfo {author} {\bibfnamefont {S.~R.}\ \bibnamefont {Nagel}},\ }\bibfield
  {title} {\bibinfo {title} {Density relaxation in a vibrated granular
  material},\ }\href@noop {} {\bibfield  {journal} {\bibinfo  {journal} {Phys
  Rev E}\ }\textbf {\bibinfo {volume} {51}},\ \bibinfo {pages} {3957} (\bibinfo
  {year} {1995})}\BibitemShut {NoStop}%
\bibitem [{\citenamefont {Nowak}\ \emph {et~al.}(1998)\citenamefont {Nowak},
  \citenamefont {Knight}, \citenamefont {Ben-Naim}, \citenamefont {Jaeger},\
  and\ \citenamefont {Nagel}}]{nowak1998density}%
  \BibitemOpen
  \bibfield  {author} {\bibinfo {author} {\bibfnamefont {E.~R.}\ \bibnamefont
  {Nowak}}, \bibinfo {author} {\bibfnamefont {J.~B.}\ \bibnamefont {Knight}},
  \bibinfo {author} {\bibfnamefont {E.}~\bibnamefont {Ben-Naim}}, \bibinfo
  {author} {\bibfnamefont {H.~M.}\ \bibnamefont {Jaeger}},\ and\ \bibinfo
  {author} {\bibfnamefont {S.~R.}\ \bibnamefont {Nagel}},\ }\bibfield  {title}
  {\bibinfo {title} {Density fluctuations in vibrated granular materials},\
  }\href@noop {} {\bibfield  {journal} {\bibinfo  {journal} {Phys Rev E}\
  }\textbf {\bibinfo {volume} {57}},\ \bibinfo {pages} {1971} (\bibinfo {year}
  {1998})}\BibitemShut {NoStop}%
\bibitem [{\citenamefont {Philippe}\ and\ \citenamefont
  {Bideau}(2002)}]{philippe2002compaction}%
  \BibitemOpen
  \bibfield  {author} {\bibinfo {author} {\bibfnamefont {P.}~\bibnamefont
  {Philippe}}\ and\ \bibinfo {author} {\bibfnamefont {D.}~\bibnamefont
  {Bideau}},\ }\bibfield  {title} {\bibinfo {title} {Compaction dynamics of a
  granular medium under vertical tapping},\ }\href@noop {} {\bibfield
  {journal} {\bibinfo  {journal} {Europhys Lett}\ }\textbf {\bibinfo {volume}
  {60}},\ \bibinfo {pages} {677} (\bibinfo {year} {2002})}\BibitemShut
  {NoStop}%
\bibitem [{\citenamefont {Philippe}\ and\ \citenamefont
  {Bideau}(2003)}]{philippe2003granular}%
  \BibitemOpen
  \bibfield  {author} {\bibinfo {author} {\bibfnamefont {P.}~\bibnamefont
  {Philippe}}\ and\ \bibinfo {author} {\bibfnamefont {D.}~\bibnamefont
  {Bideau}},\ }\bibfield  {title} {\bibinfo {title} {Granular medium under
  vertical tapping: Change of compaction and convection dynamics around the
  liftoff threshold},\ }\href@noop {} {\bibfield  {journal} {\bibinfo
  {journal} {Phys Rev Lett}\ }\textbf {\bibinfo {volume} {91}},\ \bibinfo
  {pages} {104302} (\bibinfo {year} {2003})}\BibitemShut {NoStop}%
\bibitem [{\citenamefont {Richard}\ \emph {et~al.}(2005)\citenamefont
  {Richard}, \citenamefont {Nicodemi}, \citenamefont {Delannay}, \citenamefont
  {Ribiere},\ and\ \citenamefont {Bideau}}]{richard2005slow}%
  \BibitemOpen
  \bibfield  {author} {\bibinfo {author} {\bibfnamefont {P.}~\bibnamefont
  {Richard}}, \bibinfo {author} {\bibfnamefont {M.}~\bibnamefont {Nicodemi}},
  \bibinfo {author} {\bibfnamefont {R.}~\bibnamefont {Delannay}}, \bibinfo
  {author} {\bibfnamefont {P.}~\bibnamefont {Ribiere}},\ and\ \bibinfo {author}
  {\bibfnamefont {D.}~\bibnamefont {Bideau}},\ }\bibfield  {title} {\bibinfo
  {title} {Slow relaxation and compaction of granular systems},\ }\href@noop {}
  {\bibfield  {journal} {\bibinfo  {journal} {Nature Mater}\ }\textbf {\bibinfo
  {volume} {4}},\ \bibinfo {pages} {121} (\bibinfo {year} {2005})}\BibitemShut
  {NoStop}%
\bibitem [{\citenamefont {Cundall}\ and\ \citenamefont
  {Strack}(1979)}]{cundall1979discrete}%
  \BibitemOpen
  \bibfield  {author} {\bibinfo {author} {\bibfnamefont {P.~A.}\ \bibnamefont
  {Cundall}}\ and\ \bibinfo {author} {\bibfnamefont {O.~D.}\ \bibnamefont
  {Strack}},\ }\bibfield  {title} {\bibinfo {title} {A discrete numerical model
  for granular assemblies},\ }\href@noop {} {\bibfield  {journal} {\bibinfo
  {journal} {Geotechnique}\ }\textbf {\bibinfo {volume} {29}},\ \bibinfo
  {pages} {47} (\bibinfo {year} {1979})}\BibitemShut {NoStop}%
\bibitem [{\citenamefont {Tsuji}\ \emph {et~al.}(1992)\citenamefont {Tsuji},
  \citenamefont {Tanaka},\ and\ \citenamefont {Ishida}}]{tsuji1992}%
  \BibitemOpen
  \bibfield  {author} {\bibinfo {author} {\bibfnamefont {Y.}~\bibnamefont
  {Tsuji}}, \bibinfo {author} {\bibfnamefont {T.}~\bibnamefont {Tanaka}},\ and\
  \bibinfo {author} {\bibfnamefont {T.}~\bibnamefont {Ishida}},\ }\bibfield
  {title} {\bibinfo {title} {Lagrangian numerical simulation of plug flow of
  cohesionless particles in a horizontal pipe},\ }\href@noop {} {\bibfield
  {journal} {\bibinfo  {journal} {Powder Technol.}\ }\textbf {\bibinfo {volume}
  {71}},\ \bibinfo {pages} {239} (\bibinfo {year} {1992})}\BibitemShut
  {NoStop}%
\bibitem [{\citenamefont {Hill}\ and\ \citenamefont
  {Tan}(2014)}]{hill2014segregation}%
  \BibitemOpen
  \bibfield  {author} {\bibinfo {author} {\bibfnamefont {K.~M.}\ \bibnamefont
  {Hill}}\ and\ \bibinfo {author} {\bibfnamefont {D.~S.}\ \bibnamefont {Tan}},\
  }\bibfield  {title} {\bibinfo {title} {Segregation in dense sheared flows:
  gravity, temperature gradients, and stress partitioning},\ }\href@noop {}
  {\bibfield  {journal} {\bibinfo  {journal} {J Fluid Mech}\ }\textbf {\bibinfo
  {volume} {756}},\ \bibinfo {pages} {54} (\bibinfo {year} {2014})}\BibitemShut
  {NoStop}%
\bibitem [{\citenamefont {Hill}\ and\ \citenamefont
  {Yohannes}(2011)}]{hill2011rheology}%
  \BibitemOpen
  \bibfield  {author} {\bibinfo {author} {\bibfnamefont {K.~M.}\ \bibnamefont
  {Hill}}\ and\ \bibinfo {author} {\bibfnamefont {B.}~\bibnamefont
  {Yohannes}},\ }\bibfield  {title} {\bibinfo {title} {Rheology of dense
  granular mixtures: Boundary pressures},\ }\href@noop {} {\bibfield  {journal}
  {\bibinfo  {journal} {Phys Rev Lett}\ }\textbf {\bibinfo {volume} {106}},\
  \bibinfo {pages} {058302} (\bibinfo {year} {2011})}\BibitemShut {NoStop}%
\bibitem [{\citenamefont {Trulsson}\ \emph {et~al.}(2012)\citenamefont
  {Trulsson}, \citenamefont {Andreotti},\ and\ \citenamefont
  {Claudin}}]{trulsson2012transition}%
  \BibitemOpen
  \bibfield  {author} {\bibinfo {author} {\bibfnamefont {M.}~\bibnamefont
  {Trulsson}}, \bibinfo {author} {\bibfnamefont {B.}~\bibnamefont
  {Andreotti}},\ and\ \bibinfo {author} {\bibfnamefont {P.}~\bibnamefont
  {Claudin}},\ }\bibfield  {title} {\bibinfo {title} {Transition from the
  viscous to inertial regime in dense suspensions},\ }\href@noop {} {\bibfield
  {journal} {\bibinfo  {journal} {Phys Rev Lett}\ }\textbf {\bibinfo {volume}
  {109}},\ \bibinfo {pages} {118305} (\bibinfo {year} {2012})}\BibitemShut
  {NoStop}%
\bibitem [{\citenamefont {Man}\ \emph {et~al.}(2018)\citenamefont {Man},
  \citenamefont {Feng},\ and\ \citenamefont {Hill}}]{man2018rheology}%
  \BibitemOpen
  \bibfield  {author} {\bibinfo {author} {\bibfnamefont {T.}~\bibnamefont
  {Man}}, \bibinfo {author} {\bibfnamefont {Q.}~\bibnamefont {Feng}},\ and\
  \bibinfo {author} {\bibfnamefont {K.}~\bibnamefont {Hill}},\ }\bibfield
  {title} {\bibinfo {title} {Rheology of thickly-coated granular-fluid
  systems},\ }\href@noop {} {\bibfield  {journal} {\bibinfo  {journal} {arXiv
  preprint arXiv:1812.07083}\ } (\bibinfo {year} {2018})}\BibitemShut {NoStop}%
\bibitem [{\citenamefont {Man}(2019)}]{man2019rheology}%
  \BibitemOpen
  \bibfield  {author} {\bibinfo {author} {\bibfnamefont {T.}~\bibnamefont
  {Man}},\ }\emph {\bibinfo {title} {Rheology of Granular-Fluid Systems and Its
  Application in the Compaction of Asphalt Mixtures}},\ \href@noop {} {Ph.D.
  thesis},\ \bibinfo  {school} {University of Minnesota} (\bibinfo {year}
  {2019})\BibitemShut {NoStop}%
\bibitem [{\citenamefont {Pitois}\ \emph {et~al.}(2000)\citenamefont {Pitois},
  \citenamefont {Moucheront},\ and\ \citenamefont {Chateau}}]{pitois2000}%
  \BibitemOpen
  \bibfield  {author} {\bibinfo {author} {\bibfnamefont {O.}~\bibnamefont
  {Pitois}}, \bibinfo {author} {\bibfnamefont {P.}~\bibnamefont {Moucheront}},\
  and\ \bibinfo {author} {\bibfnamefont {X.}~\bibnamefont {Chateau}},\
  }\bibfield  {title} {\bibinfo {title} {Liquid bridge between two moving
  spheres: an experimental study of viscosity effects},\ }\href@noop {}
  {\bibfield  {journal} {\bibinfo  {journal} {J. Colloid Interf. Sci.}\
  }\textbf {\bibinfo {volume} {231}},\ \bibinfo {pages} {26} (\bibinfo {year}
  {2000})}\BibitemShut {NoStop}%
\bibitem [{\citenamefont {Goldman}\ \emph {et~al.}(1967)\citenamefont
  {Goldman}, \citenamefont {Cox},\ and\ \citenamefont {Brenner}}]{goldman1967}%
  \BibitemOpen
  \bibfield  {author} {\bibinfo {author} {\bibfnamefont {A.}~\bibnamefont
  {Goldman}}, \bibinfo {author} {\bibfnamefont {R.}~\bibnamefont {Cox}},\ and\
  \bibinfo {author} {\bibfnamefont {H.}~\bibnamefont {Brenner}},\ }\bibfield
  {title} {\bibinfo {title} {Slow viscous motion of a sphere parallel to a
  plane wall¡ªii couette flow},\ }\href@noop {} {\bibfield  {journal}
  {\bibinfo  {journal} {Chem. Eng. Sci.}\ }\textbf {\bibinfo {volume} {22}},\
  \bibinfo {pages} {653} (\bibinfo {year} {1967})}\BibitemShut {NoStop}%
\bibitem [{\citenamefont {Fiscina}\ \emph {et~al.}(2010)\citenamefont
  {Fiscina}, \citenamefont {Lumay}, \citenamefont {Ludewig},\ and\
  \citenamefont {Vandewalle}}]{fiscina2010compaction}%
  \BibitemOpen
  \bibfield  {author} {\bibinfo {author} {\bibfnamefont {J.}~\bibnamefont
  {Fiscina}}, \bibinfo {author} {\bibfnamefont {G.}~\bibnamefont {Lumay}},
  \bibinfo {author} {\bibfnamefont {F.}~\bibnamefont {Ludewig}},\ and\ \bibinfo
  {author} {\bibfnamefont {N.}~\bibnamefont {Vandewalle}},\ }\bibfield  {title}
  {\bibinfo {title} {Compaction dynamics of wet granular assemblies},\
  }\href@noop {} {\bibfield  {journal} {\bibinfo  {journal} {Phys Rev Lett}\
  }\textbf {\bibinfo {volume} {105}},\ \bibinfo {pages} {048001} (\bibinfo
  {year} {2010})}\BibitemShut {NoStop}%
\bibitem [{\citenamefont {Dullien}(1992)}]{dullien1992porous}%
  \BibitemOpen
  \bibfield  {author} {\bibinfo {author} {\bibfnamefont {F.~A.}\ \bibnamefont
  {Dullien}},\ }\href@noop {} {\emph {\bibinfo {title} {Porous media: fluid
  transport and pore structure}}}\ (\bibinfo  {publisher} {Academic Press},\
  \bibinfo {year} {1992})\BibitemShut {NoStop}%
\bibitem [{\citenamefont {Jop}\ \emph {et~al.}(2006)\citenamefont {Jop},
  \citenamefont {Forterre},\ and\ \citenamefont
  {Pouliquen}}]{jop2006constitutive}%
  \BibitemOpen
  \bibfield  {author} {\bibinfo {author} {\bibfnamefont {P.}~\bibnamefont
  {Jop}}, \bibinfo {author} {\bibfnamefont {Y.}~\bibnamefont {Forterre}},\ and\
  \bibinfo {author} {\bibfnamefont {O.}~\bibnamefont {Pouliquen}},\ }\bibfield
  {title} {\bibinfo {title} {A constitutive law for dense granular flows},\
  }\href@noop {} {\bibfield  {journal} {\bibinfo  {journal} {Nature}\ }\textbf
  {\bibinfo {volume} {441}},\ \bibinfo {pages} {727} (\bibinfo {year}
  {2006})}\BibitemShut {NoStop}%
\bibitem [{\citenamefont {Boyer}\ \emph {et~al.}(2011)\citenamefont {Boyer},
  \citenamefont {Guazzelli},\ and\ \citenamefont
  {Pouliquen}}]{boyer2011unifying}%
  \BibitemOpen
  \bibfield  {author} {\bibinfo {author} {\bibfnamefont {F.}~\bibnamefont
  {Boyer}}, \bibinfo {author} {\bibfnamefont {{\'E}.}~\bibnamefont
  {Guazzelli}},\ and\ \bibinfo {author} {\bibfnamefont {O.}~\bibnamefont
  {Pouliquen}},\ }\bibfield  {title} {\bibinfo {title} {Unifying suspension and
  granular rheology},\ }\href@noop {} {\bibfield  {journal} {\bibinfo
  {journal} {Phys Rev Lett}\ }\textbf {\bibinfo {volume} {107}},\ \bibinfo
  {pages} {188301} (\bibinfo {year} {2011})}\BibitemShut {NoStop}%
\end{thebibliography}%

\end{document}